\documentclass{aa}
\usepackage{epsfig}
\begin{document}
\newcommand{\bea}{\begin{eqnarray}}    
\newcommand{\eea}{\end{eqnarray}}      
\newcommand{\be}{\begin{equation}}
\newcommand{\ee}{\end{equation}}
\newcommand{\bef}{\begin{figue}}
\newcommand{\eef}{\end{figure}}
\newcommand{\etal}{et al.}
\newcommand{\kms}{\,{\rm km}\;{\rm s}^{-1}}
\newcommand{\hubunits}{\,\kms\;{\rm Mpc}^{-1}}
\newcommand{\hmpc}{\,h^{-1}\;{\rm Mpc}}
\newcommand{\hkpc}{\,h^{-1}\;{\rm kpc}}
\newcommand{\msun}{M_\odot}
\newcommand{\K}{\,{\rm K}}
\newcommand{\cm}{{\rm cm}}
\newcommand{\cd}{{\langle n(r) \rangle_p}}
\newcommand{\Mpc}{{\rm Mpc}}
\newcommand{\kpc}{{\rm kpc}}
\newcommand{\xir}{{\xi(r)}}
\newcommand{\xrp}{{\xi(r_p,\pi)}}
\newcommand{\xsirpi}{{\xi(r_p,\pi)}}
\newcommand{\wrp}{{w_p(r_p)}}
\newcommand{\gr}{{g-r}}
\newcommand{\Navg}{N_{\rm avg}}
\newcommand{\Mmin}{M_{\rm min}}
\newcommand{\fiso}{f_{\rm iso}}
\newcommand{\Mr}{M_r}
\newcommand{\rp}{r_p}
\newcommand{\zmax}{z_{\rm max}}
\newcommand{\zmin}{z_{\rm min}}

\def\eg{{e.g.}}
\def\ie{{i.e.}}
\def\spose#1{\hbox to 0pt{#1\hss}}
\def\ltapprox{\mathrel{\spose{\lower 3pt\hbox{$\mathchar"218$}}
\raise 2.0pt\hbox{$\mathchar"13C$}}}
\def\gtapprox{\mathrel{\spose{\lower 3pt\hbox{$\mathchar"218$}}
\raise 2.0pt\hbox{$\mathchar"13E$}}}
\def\inapprox{\mathrel{\spose{\lower 3pt\hbox{$\mathchar"218$}}
\raise 2.0pt\hbox{$\mathchar"232$}}}

\title{Large scale correlations in galaxy clustering 
from the Two degree Field Galaxy Redshift Survey}

\subtitle{}

\author{N. L. Vasilyev \inst{1}, Yu. V. Baryshev \inst{1}, 
and F. Sylos Labini \inst{2,3}}

\titlerunning{Large scale correlations from the 2dFGRS}
\authorrunning{Vasilyev et al.}

\institute{Institute of Astronomy, St.Petersburg 
State University, Staryj Peterhoff, 198504,
St.Petersburg, Russia
\and``Enrico Fermi Center'', Via Panisperna 89 A, 
Compendio del Viminale, 00184 Rome, Italy
\and``Istituto dei Sistemi Complessi'' CNR, 
Via dei Taurini 19, 00185 Rome, Italy}

\date{Received / Accepted}

\abstract{We study galaxy correlations from samples extracted from the 2dFGRS
final release. Statistical properties are characterized by studying
the nearest neighbor probability density, the conditional density and
the reduced two-point correlation function. The result is that the
conditional density has a power-law behavior in redshift space
described by an exponent $\gamma=0.8 \pm 0.2$ in the interval from
about 1 Mpc/h, the average distance between nearest galaxies, up to
about 40 Mpc/h, corresponding to radius of the largest sphere
contained in the samples. These results are consistent with other
studies of the conditional density and are useful to clarify the
subtle role of finite-size effects on the determination of the
two-point correlation function in redshift and real space.
}
\maketitle

\section{Introduction}

The problem of the quantitative characterization of the large scale
galaxy clustering has been intensively discussed in the last years,
especially in relation to two new galaxy surveys: the Sloan Digital
Sky Survey (SDSS --- York et al., 2000) and the Two degree Field
Galaxy Redshift Survey (2dFGRS --- Colless et al., 2003). These data
represent a great improvement for our knowledge of the local universe:
for example the number of measured redshifts has grown of a factor ten
with respect to the surveys completed in the last two
decades. Moreover accurate redshift determinations and the multi-bands
photometry allow one a precise characterization of many parameters and
effects (e.g. K corrections) which were poorly constrained up to few
years ago. It should however be noted that for some analyses, like the
ones we discuss here, a large solid angle is also required. This is
still not the case for the present data, but, for instance, the final
release of the SDSS will provide a large contiguous angular sky region in
the very near future.

In this paper we discuss the analysis of two-point correlation
properties in the 2dFGRS sample.  Up to now these data were mainly
analyzed by studying the reduced correlation function $\xi(r)$, in
redshift and real space, and its Fourier conjugate, the power spectrum
(e.g. N`orberg et al. 2001, 2002; Tegmark, Hamilton \& 
Xu 2002;  Hawkins et al. 2003; Madgwick  et  al. 2003, 
Basilakos \& Plionis 2003, Cole et al.  2005). Recently Gaztanaga et
al. (2005) present new result for the 3-point correlation function
measured as a function of scale, luminosity and color using the 2dFGRS
sample.

In general, these statistical tools can be affected by finite-size
effects or luminosity dependent selection effects (e.g.  Gabrielli et
al., 2004) and, by using appropriate statistics, one may perform
several tests in order to disentangle different biases.  Finite size
effects can be very important for the determination of correlation
properties in the regime of large fluctuations, which should be then
clearly identified in the studies of galaxy samples.  It is in fact
well known that at small scales, observed galaxy structures are highly
irregular and present two-point power-law correlations, in the regime
of strong clustering.  However the search for the ``maximum'' size of
galaxy structures and voids, beyond which the distribution becomes
essentially uniform and fluctuations can be considered small
perturbations with respect to the average density, is still an open
problem (Tikhonov \& Makarov 2003, Hogg et al. 2005, Joyce et al. 2005
and see for a recent review Baryshev \& Teerikorpi 2005). It is
evident that from the theoretical point of view the understanding of
the statistical characteristics of these structures represents the key
element to be considered by a physical theory dealing with their
formation.

A number of statistical methods can be used to study galaxy
distribution, the main ones involve the determination of two-point
properties although the study of the distribution function, containing
information on higher order correlations, has also been found to be a
powerful method (e.g. Sivakoff \& Saslaw 2005).  The primary questions
in correlation analysis of three dimensional galaxy distributions are:
(i) what is the value of the correlation exponent and (ii) which is
the scale where the distribution becomes uniform and a {\it crossover
to homogeneity} can be clearly identified ?  Such a scale can be
defined, for example, to be the one beyond which conditional counts of
galaxies in three dimensional volumes of radius $R$ grow as
$R^3$. Recently Hogg et al. (2005), by considering the properties of a
deep and complete sample of luminous red galaxies extracted from the
SDSS survey, found that the transition from the strongly correlated
regime to the uniform one occurs at about 70 Mpc/h\footnote{Note that
we use as Hubble constant the value $H_0$=100 h $\;\;$ km/sec/Mpc
where h is $0.4\le h
\le 0.7$}, which is larger than, for example,
results in the CfA1 redshift survey where the transition was found at
about 20 Mpc/h (Davis \& Peebles, 1983; see Peebles 2001 for a recent
discussion).  Particularly, they have measured the behavior of the
conditional density in redshift space, finding that the exponent
characterizing power law correlation is about $\gamma \approx 1$
(instead of $\gamma=1.8$ as measured by Davis \& Peebles 1983) up to
20-30 Mpc/h and that this is followed by a slow crossover toward
homogeneity which is reached at about 70 Mpc/h. These results are in
good agreement with the ones presented in, e.g. Sylos Labini et
al. (1998) (see Baryshev \& Teerikorpi, 2005 for a recent review)
where the same value $\gamma
\approx 1$ was found up to 20-30 Mpc/h and where at 
larger scales, with a weaker statistics, an evidence for a
compatibility with the extension of such a behavior was found. In
addition Tikhonov, Makarov
\& Kopylov (2000) found similar results up to scales of $\sim 30$
Mpc/h, and weaker evidences for homogeneity at scales larger than
100 Mpc/h.

In this paper we present results of a correlation analysis of the
2dFGRS data studying the behavior of the conditional density and other
statistics suitable to characterize properties of distributions with
large fluctuations and control finite size effects.  In Sec.2 we
describe the procedure to construct samples which are not biased by
the luminosity selection in apparent magnitudes (the so-called volume
limited --- VL --- samples). In Sec.3 we consider the nearest neighbor
probability density for the VL samples which allows us a
characterization of small scales statistical properties. We then turn
to the study of large scale in Sec.4 where we discuss the estimation
of the conditional density and the result obtained in the VL samples.
We discuss the relation of this statistical tool with the reduced
two-point correlation function in Sec.5, where we compare our results
with previous estimations of the same statistics, focusing on finite
size effects and their implication for the interpretation of galaxy
correlations.  Finally in Sec.6 we summarize our results and discuss
its relation to other studies and we draw our main conclusions.

%%%%%%%%%%%%%%%%%%%%%%%%%%%%%%%%%%%%%%%%%%%%%%%%%%%%%%%%%%%%%%%%%%%%

\section{Volume limited subsamples}

The 2dFGRS is the largest galaxy catalog completed at the moment.  The
Final Release (Colless et al., 2003) contains more than $220$
thousands of precisely measured redshifts of the galaxies located in
two strips: about $140$ thousands in the southern galactic pole (SGP),
in a strip of $90^\circ\times 15^\circ$ and about $70$ thousands in
the strip $75^\circ\times 10^\circ$ in northern galactic pole (NGP) In
addition the survey contains $10$ thousands in the random fields which
are not used in this paper.

The median redshift of galaxies is $z\simeq0.1$ and most of the
galaxies have $z<0.3$. The $b_J$ magnitude corrected for the galactic
extinction is limited as $14.0<b_J<19.45$.

\subsection{Selection of subsamples}

To avoid the effect of the irregular edges in the angular coordinates,
due to the survey geometry, we set the following limits in  right ascension
and declination in order to get rectangular (in $\alpha, \,\delta$
coordinates) shape on the sky:
\begin{itemize}
\begin{item}
SGP:    $84^\circ\times   9^\circ$      ($-33^\circ<\delta<-24^\circ$,
$-32^\circ<\alpha<52^\circ$)
\end{item}
\begin{item}
NGP:      $60^\circ\times     6^\circ$     ($-4^\circ<\delta<2^\circ$,
$150^\circ<\alpha<210^\circ$)
\end{item}
\end{itemize}

We select galaxies in the redshift interval 
$0.01 \leq z \leq 0.3$ and with  redshift quality parameter such that 
$Q \geq 3$ in order to have high quality redshifts 
(see discussion in Hawkins et al. 2003).

We do not use a correction for the redshift completeness mask and for
the fiber collision effects. In fact, completeness varies mostly
nearby the survey edges which are excluded in our sample. We assume
that fiber collisions do not make a sensible change in the small
scales correlation properties as we set our lower cut-off to 0.5 Mpc/h
which is larger than 0.1 Mpc/h used by Hawkins et al. (2003).

To construct VL subsamples first we compute metric distances as
\begin{equation}
\label{MetricDistance}
r(z) = \frac{c}{H_0}\int_{\frac{1}{1+z}}^{1}
{\frac{dy}{y \cdot \left(\Omega_M/y+\Omega_\Lambda
\cdot y^2 \right)^{1/2}}} ~ ,
\end{equation}
where we use the standard model parameters $\Omega_M=0.3$ and
$\Omega_\Lambda=0.7$.  The absolute magnitude can be computed as
\begin{equation}\label{AbsoluteMagnitude}
  M = b_J - 5 \cdot \log_{10}\left[r(z) \cdot (1+z)\right] - K_i(z) - 25.
\end{equation}
To calculate the K-correction $K_i(z)$ (the index $i$ defines the
galaxy type) we used formulas obtained by Madgwick et al. (2002):
\begin{equation}\label{k-z}
 \begin{array}{lcl}
  K_1(z) & = 2.6z + 4.3z^2 &     \mbox{(E/S0)}\\
  K_2(z) & = 1.9z + 2.2z^2 &     \mbox{(Sa/Sb)}\\
  K_3(z) & = 1.3z + 2.0z^2 &     \mbox{(Sc/Sd)}\\
  K_4(z) & = 0.9z + 2.3z^2 &     \mbox{(Irr)}\\
  K_{avg}(z) & = 1.9z + 2.7z^2 & \mbox{(average)}\\
 \end{array}
\end{equation}
where $1,2,3,4$ indexes represent the spectral types of the galaxies
(in parenthesis in Eq.\ref{k-z}), and the average value $K_{avg}(z)$
is used for the galaxies with the undefined spectral type.

%%%%%%%%%%%%%%%%%%%%%%%%%%%%%%%%%%%%%%%%%%%%%%%%%%%%%%%%%%%%%%%%%%%%%%%%

\subsection{Definition of volume limited subsamples}

To take into account the selection effect that arises due to the
2dFGRS apparent magnitude limits $14 < b_J < 19.45$, one has to
consider two limits for the metric distance $r_{min} < r <r_{max}$ and
compute the two corresponding limits for the absolute magnitude
$M_{min}(r_{min})$ and $M_{max}(r_{max})$ which represent the lower and
the upper limit for the galaxies contained  in a VL sample. 

To this aim, we select three distance intervals (50-250 Mpc/h, 100-400
Mpc/h and 150-550 Mpc/h) and compute the corresponding absolute
magnitude limits for each of two strips.  Thus we get three VL
subsamples for the Northern hemisphere and three for Southern
hemisphere whose main parameters are presented
Table~\ref{tbl_VLSamplesProperties}. (Note that hereafter we set $h=1$
unless specified).  An example of the distance-magnitude limits for
the SGP400 sample (which indeed is the largest one considered in this
paper) is shown in Fig.~\ref{FIG1}. In Fig~\ref{FIG1b} we show
the behavior of the differential number counts $dN(r)/dr$ as a
function of distance in different sky areas for the sample
SGP400. Particularly we put limits respectively at $\delta \le
-27^{\circ}$ (c4), $\delta > -27^{\circ}$ (c5), $\alpha \ge
189^{\circ}$ (c6) and $\alpha \ge 189^{\circ}$ (c7). As an example we
report the best fit for the sample c4, which show an exponent
corresponding to a metric dimension larger ($D=3.7$) than the space
dimension. This is a purely finite-size effect corresponding to the
large fluctuations still visible at scales of order 100 Mpc/h. 
\begin{figure}
\begin{center}
\includegraphics*[angle=0, width=0.5\textwidth]{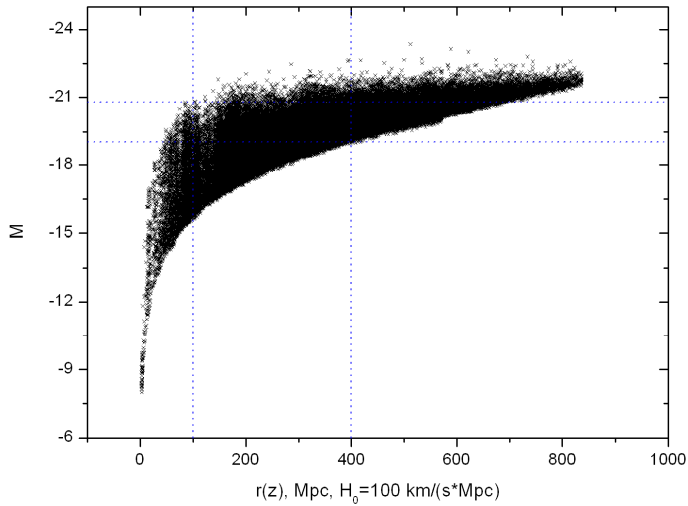}
\end{center}
\caption{The metric distance- absolute  
magnitude diagram for the SGP strip. The boundaries of the SGP400
subsample are shown.}
\label{FIG1}
\end{figure}
\begin{figure}
\begin{center}
\includegraphics*[angle=0, width=0.5\textwidth]{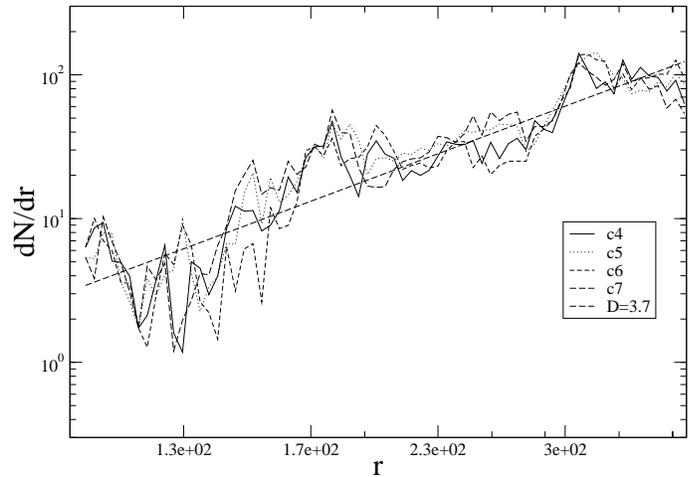}
\end{center}
\caption{ Differential number counts 
in different sky areas (defined in the text) for the SGP400 subsample.
As an example we report the best fit for the sample c4, which show an
exponent corresponding to a metric dimension larger than the space
dimension. This is a purely finite-size effect which maybe explained
by a presence of the large scale fluctuations in the studied region.}
\label{FIG1b}
\end{figure}

\begin{table}
\begin{center}
\begin{tabular}{|l|c|c|c|c|c|}
  \hline
  VL sample & $r_{min}$ & $r_{max}$ & $M_{min}$ & $M_{max}$ & $N_g$\\
  \hline
    SGP250 & 50 & 250 & -19.5 & -17.8 & 14177 \\
    SGP400 & 100 & 400 & -20.8 & -19.0 & 29373 \\
    SGP550 & 150 & 550 & -21.2 & -19.8 & 26289 \\
    NGP250 & 50 & 250 & -19.5 & -17.8 & 12474 \\
    NGP400 & 100 & 400 & -20.8 & -19.0 & 23208 \\
    NGP550 & 150 & 550 & -21.2 & -19.8 & 18030 \\
  \hline
\end{tabular}
\end{center}
\caption{Main properties 
of the obtained VL samples: $r_{min}$, $r_{max}$ are the chosen limits
for the metric distance; ${M_{min}, \,M_{max}}$ are the interval for
the absolute magnitude and $N_g$ is the resulting number of galaxies
in each sample. }
\label{tbl_VLSamplesProperties}
\end{table}

%%%%%%%%%%%%%%%%%%%%%%%%%%%%%%%%%%%%%%%%%%%%%%%%%%%%%%%%%%%%%%

%%%%%%%%%%%%%%%%%%%%%%%%%%%%%%%%%%%%%%%%%%%%%%%%%%%%%%%%%%%%%%

%%%%%%%%%%%%%%%%%%%%%%%%%%%%%%%%%%%%%%%%%%%%%%%%%%%%%%%%%%%%%%

\section{Nearest neighbor probability density}

In a stochastic point process the probability $\omega(r)dr$ that 
the nearest neighbor to a given particle lies at a distance in the
range $[r,r+dr]$ can provide a useful characterization of small scale
statistical properties. This probability density satisfies, by
definition, the condition
\begin{equation}\label{omega-norm}
\int_0^\infty{\omega(r)dr} = 1.
\end{equation}
According to its definition $\omega(r)$ can be simply estimated as
\begin{equation}\label{omegaE-r}
\omega_E(r) = N_{nn}(r)/\left(\int_0^\infty{N_{nn}(r^{'})dr^{'}}\right),
\end{equation}
where $N_{nn}(r)$ is the number of points which have their nearest
neighbors in the range $[r,r+dr]$.

The nearest neighbor probability density for a Poisson distribution
with average density $\langle{n}\rangle$, is given by (Gabrielli
et al. 2004) 
\begin{equation}\label{omega-r-Poisson}
\omega(r) = 4\pi\langle{n}\rangle{r^2}\exp
\left(-\frac{4\pi\langle{n}\rangle{r^3}}{3}\right).
\end{equation}

In Fig. \ref{FIG2} we present an example of the observed
$\omega(r)$ distribution in the VL SGP400, along with an artificial
Poisson distribution with the same number of points in the same
three-dimensional volume. Note that the probed scales here are about
$0.1 \div 10 $ Mpc/h.  

For the actual data the average distance between nearest galaxies is
smaller than for the Poisson case, and this is a clear evidence of the
presence of small scale correlations. The exact analytical behavior of
$\omega(r)$ for the general case of a power-law correlated structure
is unknown; an approximate relation for the simple case of a
anisotropic Poisson distribution, which present a radial density
profile decaying as $n_c(r) \sim r^{-\alpha}$ from its center with
exponent $\alpha$ (with $\alpha>1.5$ --- see discussion in Gabrielli
et al. 2004), is given by
\begin{equation}\label{omega-r-Fractal}
  \omega(r) = 4\pi C{r^{2-\gamma}}\exp\left(
  -\frac{4\pi C}{3-\gamma}r^{3-\gamma}\right)\;, 
\end{equation}
where $\gamma=3-2\alpha$.   This is found to be a good approximation in
the actual data (see Fig.\ref{FIG1}).  In
Tab.\ref{tbl_VLSamplesScales} (see below) we report the estimation of
the average distance between nearest neighbors  (defined as $r_{sep} = \int r
\omega(r) dr$) in the different samples (Note that similar values 
have been found by Peebles 2001).
\begin{figure}
\begin{center}
\includegraphics[angle=0, width=0.5\textwidth]{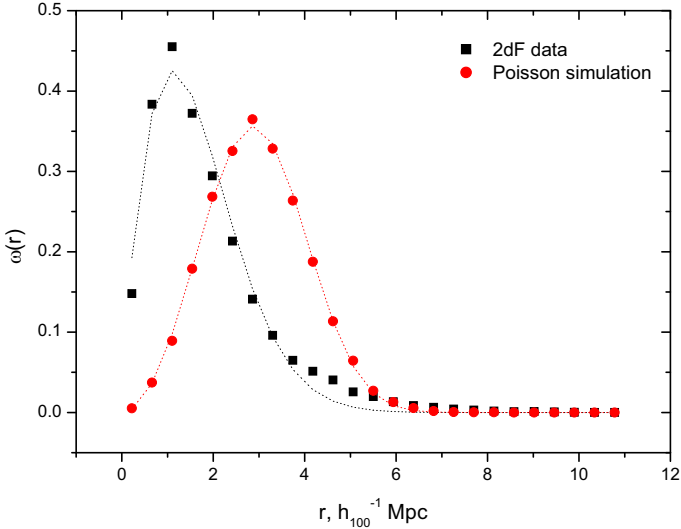}
\end{center}
\caption{Nearest neighbor probability 
density for the SGP400 sample data (squares) and for a Poisson
simulation (circles) in the same volume. The dotted lines are the best
fit respectively for the anisotropic Poisson distribution
(Eq.\ref{omega-r-Fractal} with $\gamma=1.2$) and for the Poisson case
(Eq.\ref{omega-r-Poisson}). }
\label{FIG2}
\end{figure}
%
%%%%%%%%%%%%%%%%%%%%%%%%%%%%%%%%%%%%%%%%%%%%%%%%%%%%%%%%%%%%%%%%%%%%%%%%%%%%%%%%%%%%5

\section{Estimation of correlations: the  conditional density}

In general, in a distribution of points with large fluctuations at
some scales, one may determinate two-point correlations through the
estimation of the conditional density (see discussion in Gabrielli et
al. (2004)). We first briefly summarize the main properties of this
statistical tool stressing the finite size effects and statistical
errors which may enter into the estimators.  Then we apply it to the
case of the VL samples extracted from the 2dFGRS, as discussed in the
previous section.

%%%%%%%%%%%%%%%%%%%%%%%%%%%%%%%%%%%%%%%%%%%%%%%%%%%%%%%%%%%%%%%%%%%%%%%%%%%%%%%%%

\subsection{Conditional density in spheres $\Gamma^*(r)$}

The conditional density in spheres $\Gamma^*(r)$ is defined for an
ensemble of realizations of a given point process, as
\begin{equation}\label{Gamma*-r}
 \Gamma^*(r) 
 = \frac{\langle{N(r)}\rangle_{P}}{\|C(r)\|}. 
\end{equation}
This quantity measures the average number of points
$\langle{N(r)}\rangle_{P}$ contained in a sphere of volume
$\|C(r)\|=\frac{4}{3}\pi{r}^{3}$ with the condition that the center of
the sphere lies on an occupied point of the distribution (and
$\langle{...}\rangle_{P}$ denotes the conditional ensemble average).

Such a quantity can be estimated in a finite sample by a volume
average (supposing stationarity of the point distribution)
\begin{equation}\label{Gamma*E-r}
 \Gamma^*_E(r) = \frac{\overline{N(r)}_P}{\|C(r)\|}
 = \frac{1}{N_c(r)} \sum_{i=1}^{N_c(r)}{\frac{N_i(r)}{\|C(r)\|}},
\end{equation}
where $N_c(r)$ --- the number of points (centers) with balls fully
contained in the sample volume, $\overline{(...)}_{P}$ means averaging
by the sample points.

Given a sample of arbitrary geometry and a scale $r$ at which
correlations are measured, only a subsample of the points contained in
it will satisfy the following requirement: when chosen as center of a
sphere of radius $r$, the sphere is fully contained in the sample
volume. When the average in Eq.\ref{Gamma*E-r} is made over such a
subsample one is considering the full-shell estimator of the
conditional density. Note that the number of center $N_c(r)$ is a
function of the scale $r$ at which correlations are estimated.  In
fact for scales much smaller than the radius $r_s^m$ of the largest
sphere fully contained in the sample volume, almost all points will
contribute to the average, while at scales comparable to the sample
size only those points lying in the center of the sample volume will
contribute. Thus finite-size effect can be important when one
considers the largest available scales: in this situation one cannot
make a full volume average and systematic effect, due to large
fluctuations, can be important in the determination of such a
statistics.

The scale $r_s^m$ will in general be very different from the scales
$r_{min}$ and $r_{max}$ characterizing a VL sample, as it depends
crucially on the sample solid angle. On the other hand the minimal
scale $r_{sep}$ up to which correlations can be measured, is given by
the average distance between neighbor galaxies: clearly for
$r<r_{sep}$ discrete shot-noise dominates estimations of any
statistical quantity. Thus we will explicitly compute the scales
$r_{sep}$ and $r_s^m$ for the VL considered in what follows (see
Tab.\ref{tbl_VLSamplesScales}).

%%%%%%%%%%%%%%%%%%%%%%%%%%%%%%%%%%%%%%%%%%%%%%%%%%%%%%%%%%%%%%%%%%%%%%

\subsection{Conditional density in shells $\Gamma(r)$}

The conditional density in spherical shells is defined as
\begin{equation}\label{Gamma-r}
 \Gamma(r) = \frac{\langle{N(r,\Delta{r})}\rangle_P}{\|C(r,\Delta{r})\|},
\end{equation}
where $\langle{N(r,\Delta{r})}\rangle_P$ represents the ensemble
average number of points in a sphere of radius $r$ and thickness
$\Delta{r}$, of volume
$\|C(r,\Delta{r})\|=\frac{4}{3}\pi[(r+\Delta{r})^{3}-r^3]$, around a
point of distribution (and thus this is a conditional ensemble average
$\langle{...}\rangle_{P}$ as in the previous case).  Note that one can
also write Eq.\ref{Gamma-r} as
\begin{equation}\label{Gamma-r2}\Gamma(r) =
\frac{\langle{n(r)n(0)}\rangle}{\langle{n(0)}\rangle}
\end{equation}
where $\langle ... \rangle$ represents the (unconditional) ensemble
average and $n(r)$ is the microscopic number density.

The conditional density in shells can be estimated in a finite sample
by the following volume average
\begin{equation}\label{GammaE-r}
 \Gamma_E(r) = \frac{\overline{{N(r,\Delta{r})}}_P}
  {\|C(r,\Delta{r})\|} = \frac{1}{N_c(r+\Delta{r})}
 \sum_{i=1}^{N_c(r+\Delta{r})}{\frac{{N_i}(r,\Delta{r})}
  {\|C(r,\Delta{r})\|}},
\end{equation}
where we consider again only the full-shell estimator, i.e. for which
$N_c(r+\Delta{r})$ represents the number of points (centers) contained
in spherical shells fully contained in the sample volume. Analogously
to the case of $\Gamma^*(r)$ particular care should be used to
determine the scales $r_{sep}$ and $r_s^m$.

 It is instructive to notice that for the case where the distributions
 has power-law correlations and strong fluctuations (e.g. a fractal
 structure) then the conditional density in spheres behaves (in the
 ensemble average) as
\begin{equation}\label{Gamma*-r-fractal}
  \Gamma^*(r) = \frac{3B}{4\pi}~r^{-\gamma}\,,
\end{equation}
while conditional density in shells has the form
\begin{equation}\label{Gamma-r-fractal}
  \Gamma(r) = \frac{(3-\gamma)B}{4\pi}~r^{-\gamma}.
\end{equation}
where $\gamma$ is the correlation exponent (in the case of a fractal
$D=3-\gamma$ is the fractal dimension) and $B$ is a lower cut-off
related to the smaller scale where correlation can be measured in a
finite sample (i.e. to $r_{sep}$ previously defined). 

%%%%%%%%%%%%%%%%%%%%%%%%%%%%%%%%%%%%%%%%%%%%%%%%%%%%%%%%%%%%%%%%%%%%%%%%%%%%%%%%%%%%%%%%%

\subsection{Application to 2dFGRS data}

In Tab.\ref{tbl_VLSamplesScales} we show, for the different VL samples
considered, the lower and upper cut-off, previously discussed, between
which we have estimated $\Gamma(r)$ and $\Gamma^*(r)$. Note that we
have generated a Poisson distribution, for each VL sample, with the
same number of points and in the same three dimensional volume in
order to estimate the same statistical quantities in a distribution
without correlation at all.  This provide us with a useful way to test
our analysis with the simplest distribution with known
properties. Note also that all our estimates have been done in {\it
redshift space}: the relation  with real space properties will be
discussed in Sec.6. 
\begin{table}
\begin{center}
\begin{tabular}{|l|c|c|c|}
  \hline
  VL sample & $r_{sep}$, 2dFGRS & $r_{sep}$, Poisson & $r_{s}^{m}$ \\
  \hline
    NGP250 & 1.3 & 2.0 & 12.4 \\
    NGP400 & 1.7 & 2.6 & 19.9 \\
    NGP550 & 2.7 & 3.9 & 27.4 \\
    SGP250 & 1.5 & 2.4 & 18.2 \\
    SGP400 & 1.9 & 3.0 & 29.1 \\
    SGP550 & 2.8 & 4.2 & 40.0 \\
  \hline
\end{tabular}
\end{center}
\caption{Characteristic 
scales of the VL samples: $r_{sep}$ is the average separation distance
between nearest neighbor galaxies (in 2dFGRS and Poisson distribution
within the same volume and for the same number of galaxies),
$r_{s}^{m}$ is the maximum sphere completely contained in the
sample. All distances are in Mpc ($H_0=100$ km/sec/Mpc).} 
\label{tbl_VLSamplesScales}
\end{table}

Fig.\ref{fig_GammaInt}
\begin{figure}
\begin{center}
\includegraphics*[angle=0, width=0.5\textwidth]{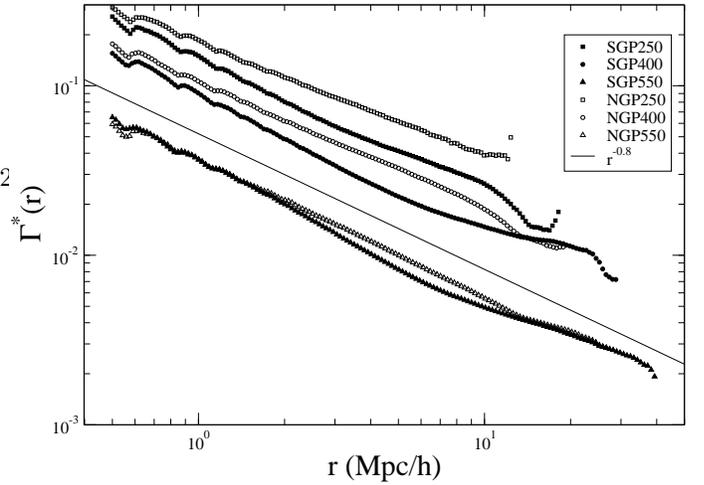} 
\end{center}
\caption{Estimation of the
conditional density in spheres in the six VL samples considered (different symbols
correspond to different VL samples --- see labels).  The  reference
line has a power-law behavior with  slope $\gamma=0.8$. }
\label{fig_GammaInt}
\end{figure}
 shows the behavior of the estimation of the
conditional density in spheres in the six VL samples considered. It is
interesting to note that  samples with same luminosity and distance
cuts in the NGP and SGP show approximately the same behavior. However
a difference in the amplitude is present for all but the largest
sample. The amplitude of $\Gamma^*(r)$ is related to the luminosity
function in the following way. 

In general one has that the joint conditional probability of finding a
galaxy of luminosity $L$ at distance $\vec{r}$ from another galaxy,
i.e.  the (ensemble) conditional average number of galaxies with
luminosity in the range $[L,L+dL]$ and in the volume element $d^3r$ at
distance $r$ from an observer located on a galaxy is given by $\langle 
\nu(L,\vec{r}) \rangle_p d^3rdL$. One can then assume that
\be   
\label{gal2}   
\langle \nu(L,\vec{r}) \rangle_p = 
\phi(L) \times \Gamma(r) \; , 
\ee   
where $\Gamma(r)$ is the {\it average conditional density} and
$\phi(L)$ is the {\it luminosity function} such that $\phi(L)dL$ gives
the probability that a randomly chosen galaxy has luminosity in the
range $[L,L+dL]$. By writing Eq.\ref{gal2} as a product of the
conditional space density for the luminosity function, one has
implicitly assumed that galaxy positions are independent of galaxy
luminosity. Thus from Eq.\ref{gal2} it follows that the amplitude of
$\Gamma^*(r)$ in a VL sample is given by an integral of the luminosity
function over the range of absolute luminosity covered by the sample
multiplied by the conditional density for all galaxies. Amplitude
variations in the same VL samples in the NGP and SGP can be due to
large local fluctuations which are not averaged out by the volume
average. Thus these differences can be probably ascribed to finite
size effects. The fact that in the deepest VL samples (i.e. the ones
cut at 550 Mpc/h), where the volume is the largest, the conditional
density does not show significant differences between the two
hemispheres supports the finite-size interpretation.

If one fits the behavior of the estimated $\Gamma^*(r)$ with
a power-law function of the type $ Br^{-\gamma}$ one finds that
$\gamma=0.8 \pm 0.2$. In Fig.\ref{fig_GammaInt_Normalized} 
\begin{figure}
\begin{center}
\includegraphics*[angle=0, width=0.5\textwidth]{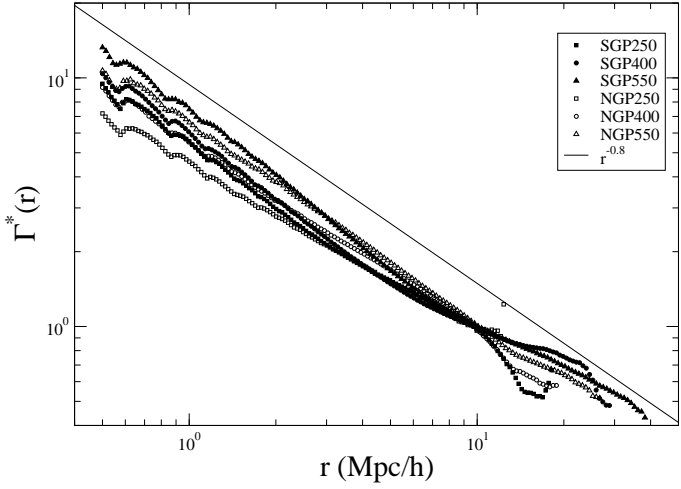}
\end{center}
\caption{Conditional density is spheres normalized  to the value at 10 Mpc/h
In this way it is apparent the fact that the slope variates in the
different samples.  The  reference
line has a power-law behavior with  slope $\gamma=0.8$. }
\label{fig_GammaInt_Normalized}
\end{figure}
we have normalized
the conditional density is spheres to the value at 10 Mpc/h. In this way it is 
apparent the fact that the slope variates in the different samples:
the variation is of about 0.1. The formal statistical error
for the determination of $\Gamma^*(r)$ at each scale can be simply derived 
from the dispersion of the average 
\be
\Sigma^2(r) = \frac{1}{N} \sum_{i=1}^{N-1} 
\frac{\left( \Gamma_i^*(r) - \Gamma^*(r) \right)^2}
{N-1} 
\ee
where $\Gamma_i^*(r)$ represents the determination from the $i^{th}$
point. The corresponding error bars are too small to be
plotted. However one should notice that at large scales (usually the
last few points) estimators of $\Gamma^*(r)$ have large scatterings
because of the small number of points contributing to the
average. Moreover in this estimation one cannot take into account
systematic variations due to the fact that the volume average cannot
be performed at large scales (see discussion in Joyce, Montuori \&
Sylos Labini 1999). For these reasons the behavior for scale larger
than $\sim 20$ Mpc/h is affected by large un-averaged fluctuations.

We show in Fig.\ref{fig_GammaDiff} the behavior of the conditional
density in shells and in Fig.\ref{fig_GammaDiff_Normalized} the
conditional density in shells normalized to the value at 10 Mpc/h. It
is clear that these estimations are more affected by statistical
noise. An important parameter in this respect is represented by the
shell thickness which we take constant in logarithmic scale.
\begin{figure}
\begin{center}
\includegraphics*[angle=0, width=0.5\textwidth]{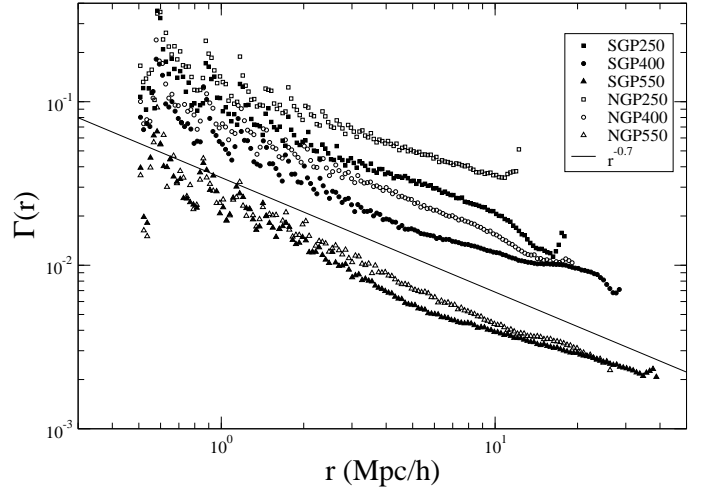}
\end{center}
\caption{Estimation of the conditional density in shells
for the different VL samples considered. The  reference
line has a power-law behavior with  slope $\gamma=0.7$. 
} \label{fig_GammaDiff}
\end{figure}
\begin{figure}
\begin{center}
\includegraphics*[angle=0, width=0.5\textwidth]{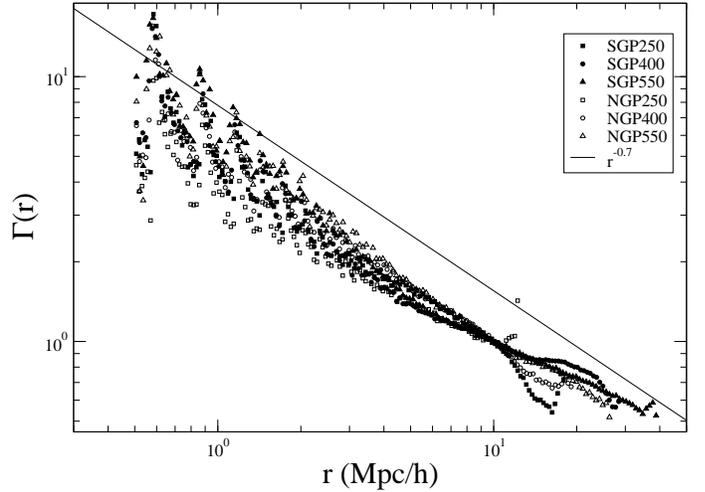}
\end{center}
\caption{Estimation of the conditional density in shells normalized
to the value at 10 Mpc/h for the different VL samples considered. The
reference line has a power-law behavior with slope $\gamma=0.7$.  }
\label{fig_GammaDiff_Normalized}
\end{figure}
In this case the average slope is $\gamma=0.8 \pm 0.2$ up to 30 Mpc/h.

\section{Estimation of  the reduced two-point correlation function} 

The reduced two-point correlation function $\xi(r)$ for a stochastic
point process is defined (see e.g. Peebles, 1980)  as
\begin{equation}
\label{xi-r}
 \xi(r) = \frac{\langle{n(r)n(0)}\rangle}{\langle{n(0)}\rangle^2}-1 =
 \frac{\Gamma(r)}{\langle{n}\rangle}-1,
\end{equation}
where $\langle{...}\rangle$ indicates the ensemble average and
$\langle{n}\rangle$ --- is the ensemble average number density. The
last equality follows from the very definition of the conditional
density (see Eq.\ref{Gamma-r}).

There are several estimators of $\xi(r)$ and we refer to Kerscher,
Szapudi \& Szalay (2000) and Gabrielli et al. (2004) for a detailed
discussion of the different ones used in the literature. One may
consider, for example, the Landy \& Szalay (1993) (LS) estimator that is
the most widespread in modern studies of correlation function for
large scale structures because it is the minimal variance estimator
for a Poisson distribution.  This can be written as (Kerscher, Szapudi
\& Szalay al. 2000):
\begin{equation}\label{xiLS-r}
 \xi_{LS}(r) =
\frac{N_R(N_R-1)}{N_D(N_D-1)}\frac{DD(r)}
{RR(r)}-2\frac{N_R-1}{N_D}\frac{DR(r)}{RR(r)}+1,
\end{equation}
where $N_D$ --- the number of data (sample) points; $N_R$ --- the
number of random points homogeneously distributed in the sample
geometry; $DD(r)$ is the number data-data pairs, $DR(r)$ ---
data-random pairs and $RR(r)$ --- random-random pairs respectively.
  Note that in the artificial random catalogs
generated for the estimation of Eq.\ref{xiLS-r}, we have used 
a number points in the range 4.5-9$10^4$. 
However the LS estimator can be biased by finite-size effects
in the case of strongly correlated distributions as we discuss 
in what follows:we have tested that also for the estimator
introduced by Davis \& Peebles (1983) the situation is 
substantially the same. 

Analogously to the full-shell estimator of the conditional
density, one may define the following (full-shell)
estimator of $\xi(r)$ which can be induced directly from Eq.\ref{xi-r}
\be
\label{xi_fs} 
\xi_{FS}(r)= \frac{\Gamma_E(r)}{\Gamma^*_E(r_s^m)} -1 \;.
\ee
where $\Gamma_E(r)$ is the estimator of the conditional density in
shells and $\Gamma^*_E(r_s^m)$ is the estimator of the conditional
density in spheres at the scale of the sample $r_s^m$.  Although the
latter quantity is not, in general, computed through an average
because only a single point may contribute at such large scales, this
estimator, when the properties of the distribution are unknown and
likely to be characterized by strong fluctuations, has several
advantages with respect to the LS (or also the one used by Davis \&
Peebles, 1983). 

We notice that by using the full-shell estimator we are able to make a
very conservative measurement of the two-point correlation
function. In fact, for example, one does not need to make estimations
of correlations on scales larger than $r_s^m$ which require use of
weighing schemes and special treatment of boundary conditions.  The
main point is however that the estimation of the sample density is
performed on ``local'' scales, i.e. much smaller than the global scale
of the sample.  In addition Eq.\ref{xi_fs} satisfies the simple
constraint
\be
\label{intcost} 
\int_0^{r_s^m} \xi_{FS}(r) r^2 dr = 0
\ee
which is the so-called ``integral constraint''. Any estimator of
$\xi(r)$ must satisfy a similar condition which comes from the fact
that the average density has been estimated from the given sample.
(Note that we do not use any additional correction to take into
account for this particular effect: in the case of the full-shell
estimator the integral constraint has a clear effect given by
Eq.\ref{intcost}. For the case of the LS estimator we have not used
any correction to take into account for this constraint.)  It is
however clear that Eq.\ref{intcost} gives us a simple way of
controlling this offset, which is not the case for another estimator.
It is important to stress that the estimation of the sample average is
subjected to large fluctuations because its determination does not
involve any average. Such fluctuations will substantially alter the
amplitude of the reduced correlation function as we discuss below:
this is a good reason to measure statistical quantities, like
$\Gamma(r)$, which are not affected by such fluctuations.

The behavior  of $\xi_{FS}(r)$ is  presented on Fig. \ref{fig_xiFS}.
\begin{figure}
\begin{center}
\includegraphics*[angle=0, width=0.5\textwidth]{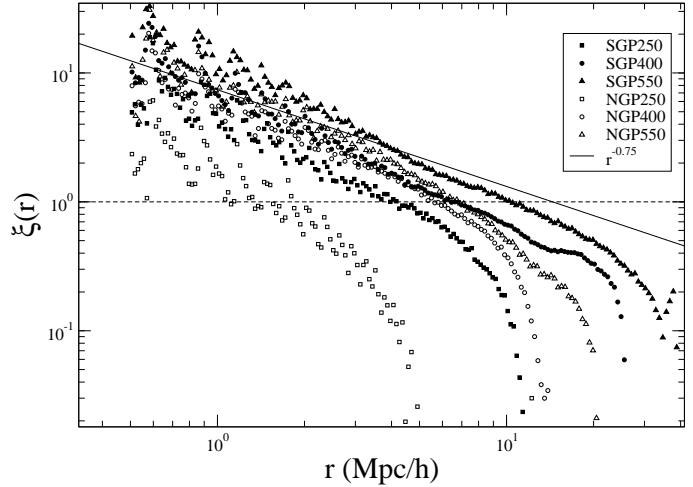}
\end{center}
\caption{Estimation of the two-point reduced correlation 
function in the different VL samples considered by using the full
shell estimator. The reference line has a power-law behavior with
slope $\gamma=0.75$.  }
\label{fig_xiFS}
\end{figure}
We note two main properties: the first one is that the amplitude of
$\xi(r)$ changes in different samples and the second is that the
exponent in the strongly clustered regime (i.e. $\xi(r) \gg 1$) is
about $\gamma=0.75$. Both results are in {\it qualitative} agreement
with other analysis of the same samples. For example Hawkins et
al. (2003) found that in the full magnitude limited sample, the 
redshift space value of the correlation exponent is $\gamma=0.75$ in
the range [0.1,4] Mpc/h and then $\gamma=1.75$ in the range [4,10]
Mpc/h (see their Fig.6).  This is for example what we find in the
SGP250 sample as shown in Fig.\ref{fig_xiFS2}. It is worth
noticing that the slopes measured in different VL sample may variate
as it is shown, for example, by the SGP400 in Fig.\ref{fig_xiFS3}.
\begin{figure}
\begin{center}
\includegraphics*[angle=0, width=0.5\textwidth]{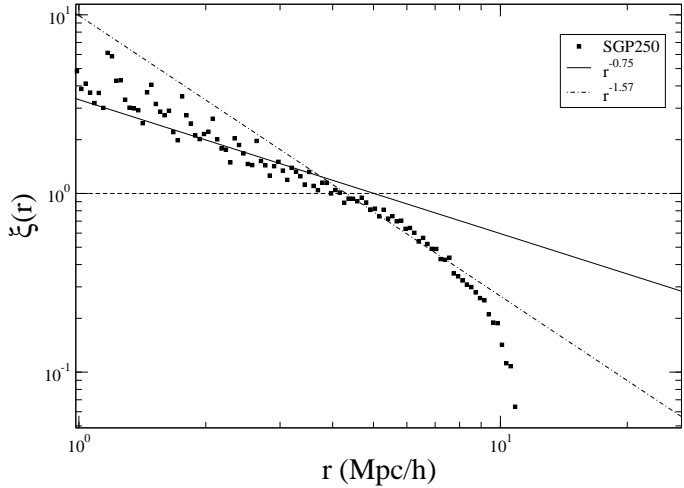}
\end{center}
\caption{Estimation of the two-point reduced correlation 
function in the VL sample SGP250, by using the full shell
estimator. The reference lines have a power-law behavior with slope
$\gamma=0.75$ and $\gamma=1.57$ respectively.}
\label{fig_xiFS2}
\end{figure}
\begin{figure}
\begin{center}
\includegraphics*[angle=0, width=0.5\textwidth]{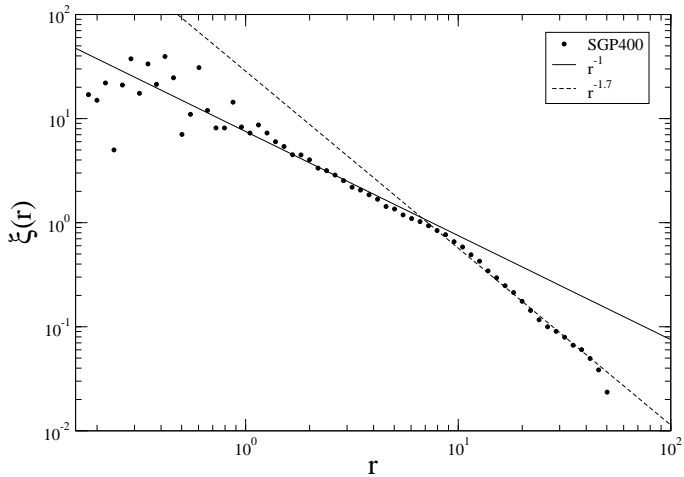}
\end{center}
\caption{Estimation of the two-point reduced correlation 
function in the VL sample SGP400, by using the full shell
estimator. The reference lines have a power-law behavior with slope
$\gamma=1.0$ and $\gamma=1.7$ respectively.}
\label{fig_xiFS3}
\end{figure}
It is interesting to note that in other surveys different values of
$\gamma$, in redshift space, have been found: for example in the CfA1
catalog $\gamma=1.8$ in the range [0.1,5] Mpc/h (Davis \& Peebles
1983).  As discussed below we ascribe this change of slope, as the
variation of the amplitude of $\xi(r)$, to a finite size effect.  For
this reason, while the qualitative behavior of the variation of the
amplitude and exponent of $\xi(r)$ is similar to many other
estimations (e.g. Hawkins et al., 2003; Norberg et al. 2001), the
quantitative comparison depends on the sample size and, most
importantly, on the fluctuations which affect the determination of the
sample density. As these fluctuations can be large and dependent on
the specific sample considered, it is difficult to make a more
quantitative comparison between our and other results.

It is also very interesting to note that the zero-crossing scale
$r_{zc}$ of $\xi(r)$, shown by a sharp decay at the scale $r_{zc}$ of
$\xi_E(r)$ in a log-log plot, depends on the sample size. This result
can be again explained as finite size effect introduced by
Eq.\ref{intcost}. This is an important feature
especially in the comparison between observations and numerical
N-body simulations (see Sylos Labini 2005 for more detail).

Concerning the amplitude, we note that Norberg et al. (2001) found a
similar variation of the redshift-space $\xi(r)$. This is consistent 
with the results discussed here. The difference lies in the
way these results are interpreted. In fact, while Norberg et
al. (2001) ascribe the different amplitudes to different selections in
luminosity (or spectral type, or colors, etc.), we discuss below that,
given the behavior of the conditional density, such variations can be
easily explained as a finite size effect.

\subsection{The role of finite size effects in redshift space} 

In order to directly show the importance of finite size effects, and
illustrate their role in a specific example, we have considered the
sample SGP400 and constructed some different subsamples. In all cases
the other boundaries in $\alpha,\delta,r$ remain the same as for the
original sample while an additional cut has been imposed.  The sample
C1 is cut at $r\le 250$ Mpc/h, C2 at $r\le 300$ Mpc/h, C3 at $r\le
350$ Mpc/h, C4 at $\delta
\le -0.5$ radiant, C5 at $\delta \ge -0.5$ radiant, C6 at $\alpha \ge
3.3$ radiant, C7 at $\alpha \le 3.3$ radiant and C8 at $r\le 315$
Mpc/h. Note that in these subsamples the lower cut-off remains
the same as for the full SGP400, while the upper cut-off changes:
in what follows we focus on how the finite size effect
at large scales influence the amplitude of the $\xi$-function.
The results obtained by the Landy-Szalay estimator
(Eq.\ref{xiLS-r}) are shown in Fig.\ref{FIG9}
\begin{figure}
\begin{center}
\includegraphics*[angle=0, width=0.5\textwidth]{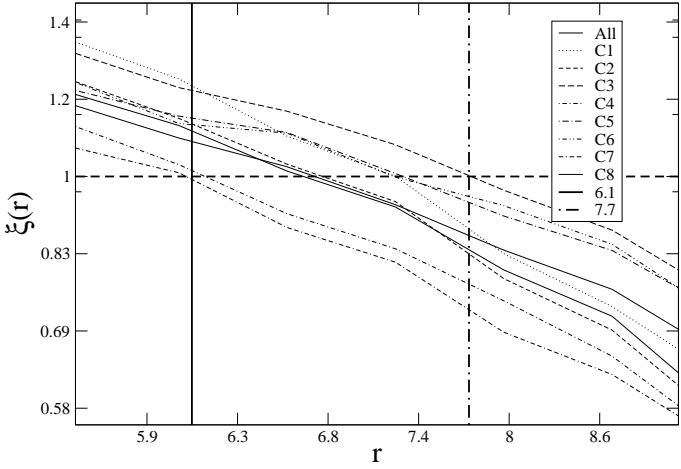}
\end{center}
\caption{Estimation of the two-point reduced correlation 
function in the different subsamples of the SGP400 VL sample by the
Landy-Szalay estimator (Eq.\ref{xiLS-r}). The length-scale $r_0$ varies
from 6.1 Mpc/h to 7.7 Mpc/h in the different samples.}
\label{FIG9}
\end{figure}
One may note that the amplitude of $\xi_E(r)$ varies in the different
subsamples. Note that we refer to the amplitude variation of
$\xi(r)$ as shown by Fig.\ref{FIG9} without making a detailed analysis
of the power-law exponent and the corresponding amplitude). The reason
for this choice lies in the insignificant values of formal statistical
errors along with large systematical errors (especially at large
scales) due to the finite volume and single realization.  Instead of
performing precise estimation of $r_{0}$ and $\gamma$ we simply
demonstrate the general behavior of $\xi$-function.  This variation is
due to fluctuations in the large scale distribution of galaxies and
thus they are volume dependent effects. Therefore the amplitude of
$\xi(r)$ is affected by finite-size effects as long as the
distribution has not been found to have relaxed to an uniform
system. From the one hand the Landy-Szalay estimator uses a sample
density computed on the global sample scale, thus introducing a
mixture of large scales and small scales properties in the measure of
correlations. From the other hand, although the sample depth is of
order of hundreds Mpc/h, finite size effects, related to the presence
of large scale structures can be still important. The use of the
conditional density avoids both these problems.

In order to explain the amplitude and slope variation observed by the
estimation of two-properties by $\xi(r)$ we introduce a simple toy
model which may capture the main element of the problem. However one
may repeat the following argument for any distribution, and thus for
any functional behavior of the conditional density, one finds in the
data. The point is that in the regime of strong clustering, evidenced
by the range of scales where $\Gamma(r)$  has not
reached a clear flattening behavior,  the determination of $\xi(r)$ and
thus of the average density, is sample size dependent.

If the conditional density has a power-law behavior up to the size
$r_s^m$ of the type
\be
\label{gamma_model}
\Gamma(r) = B r^{-\gamma}
\ee
with $0<\gamma<3$ then the estimation of the sample average 
through the conditional density in spheres is 
\be
\label{gamma_model2} 
\Gamma^*(r_s^m) = 
\frac{3}{4\pi (r_s^m)^3} \int_0^{r_s^m} \Gamma(r) 4\pi r^2 dr = 
\frac{3 B}{3-\gamma} (r_s^m)^{-\gamma} \;.
\ee
Thus from Eq.\ref{xi_fs} we find that 
\begin{equation}\label{xiE-r-Fractal}
 \xi_{FS}(r) = 
\frac{3-\gamma}{3}~\left(\frac{r}{r_s^m}\right)^{-\gamma} - 1 \;. 
\end{equation}
One may note that Eq.\ref{xiE-r-Fractal} easily takes into account
both the amplitude variation in samples of different size, and the
change of the slope as a function of scale (due to the different
regime of strong correlation where the fit with a power-law is
possible).  From Eq.\ref{xiE-r-Fractal} one may note that the slope
depends on scale in a continuous way: for example at $r=r_0$ such that
$ \xi_{FS}(r_0)=1$ one easily derives that the local slope becomes $2
\gamma$ (see Fig.\ref{fig_xiFS2}). 
In fact, Hawkins et al. (2003) fitted the slope around the scale $r_0$
in the different samples (see their Tab.1) with the consistent result
that the slope is $1.6$.

Moreover we would like to remark the crucial point that
$\Gamma^*_E(r_s^m)$ can differ from Eq.\ref{gamma_model} in a single
sample determination: while the latter is the expectation value for
the ensemble average quantity, the former quantity is subjected to
large finite size fluctuations. This implies that the scaling of the
amplitude of $\xi_{FS}(r)$ does not hold precisely in a single
measurement, while this is the expectation in an ensemble of
realizations (which is not possible to obtain in the analysis of a
single sample).

\subsection{The role of finite size effects in real space} 

Concerning the real space properties, we have not directly measured
them here. However we may notice that the same finite-size effects
which perturb the redshift space reduced two-point correlation
function may affect the projected one (usually called $\omega(r_p)$
--- see e.g. Davis \& Peebles 1983). In general, one may relate the
real space $\xi_{RS}(R)$ to the projected $\omega(r_p)$, where $r_p$
represents the projection of the redshift space distance on a
direction perpendicular to the line of sight, through the following
equation
\be
\label{omega1}
\omega(r_p) = 2 \int_{r_p}^{\infty} 
\frac{\xi_{RS}(y)y}{\sqrt{y^2-r_p^2}} dy\;.
\ee
Let us now consider the following situation: if the real space
conditional density has the behavior $\Gamma_{RS}(R)=AR^{-\gamma}$
then we can repeat the argument which yields to Eq.\ref{xiE-r-Fractal}
with the result that
\be
\label{omega2}
\xi_{RS}(R) = \frac{3-\gamma}{3} 
\left(\frac{R}{r_s^m}\right)^{-\gamma}-1
\ee
where $r_s^m$ is the sample depth, as discussed. Thus the real space
$\xi_{RS}(R)$ shows the same finite-size effects present in the
redshift space correlation function previously discussed. If
$\xi_{RS}(R)$ has a pure power-law behavior with $\gamma>1$ then 
from  Eq.\ref{omega1} one gets 
\be 
\label{omega3}
\omega(r_p) \sim r_p^{1-\gamma} \;.
\ee
 In the present situation this is not the case, because the second
 term in Eq.\ref{omega2} gives an infinite contribution when
 integrated over all space. In practice however one truncates the
 integral to scales of order of $r_s^m$ and one expects to recover
 Eq.\ref{omega3} only at small enough scales. Thus
\be
\label{omega4}
\omega(r_p) = 2 \int_{r_p}^{r_s^m} \frac{\xi_{RS}(y)y}{\sqrt{y^2-r_p^2}} dy \;.
\ee
The finite size effect introduced by the cut-off $r_s^m$ may well take
into account the observed shape of $\omega(r_p)$. For example for
$\gamma=0.8$ we get from Eq.\ref{omega2} and Eq.\ref{omega4} the
behavior shown in Fig.\ref{omega}, which is very similar to the one
measured by Hawkins et al. (2003).  Hence, while in this example
$\gamma<1$ we get that $\omega(r_p) \sim r_p^{-1.5}$ in some finite
range of scales: this is a finite size effect similar to the one
varying the estimation of the exponent of the redshift space correlation
function.  Note that a similar finite-size effect may be present in
the measurement of the angular two-point correlation function (see
e.g., Montuori \& Sylos Labini 1997).
\begin{figure}
\begin{center}
\includegraphics*[angle=0, width=0.5\textwidth]{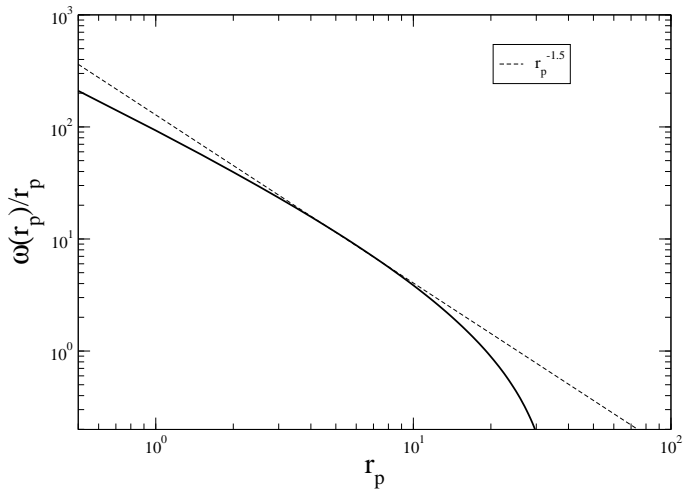}
\end{center}
\caption{Behavior of $\omega(r_p)/r_p$ computed numerically 
from Eq.\ref{omega2} and Eq.\ref{omega4} with $\gamma=0.8$ and
$r_s^m=70$ Mpc/h. The dashed line has a slope of $-1.5$. 
}
\label{omega}
\end{figure}
While we are clearly not able to make a definitive statement about
whether the behavior of $\omega(r_p)$ found by, for instance, Hawkins
et al. (2003), is perturbed by systematic biases, our analysis shows
that one needs to consider finite size effects explicitly also for the
computation of the real space properties.

%%%%%%%%%%%%%%%%%%%%%%%%%%%%%%%%%%%%%%%%%%%%%%%%%%%%%%%%%%%%%%%%%%

\section{Discussion}

We have studied redshift space correlation properties of six volume
limited samples extracted from the 2dFGRS. We have considered several
statistical properties. Particularly the characterization of
small-scale properties through the nearest neighbor probability
density allow us the determination of the smallest scale up to which
correlations properties can be studied in a robust way. In fact, at
scales smaller than the average distance between nearest neighbors,
typically in the range of few Mpc/h (see
Tab.\ref{tbl_VLSamplesScales}), discrete shot noise dominates the
measurements leading to deviations from a power-law behavior. Whether
the result of Zehavi et al. (2004), who found departures from a power
law behavior in the galaxy correlation function of some samples of the
SDSS catalog can be interpreted in this way, i.e. as dominated by
nearest-neighbor correlations, is an open question, as they did not
mention what is the average distance between nearest neighbors in
their sample, and then they performed the analysis in real space
instead of redshift space as we do here.

For the conditional average density we find that it is characterized
fairly well by a power-law behavior in the range between 0.5 and 40
Mpc/h, where the exponent is $\gamma= 0.8\pm 0.2$. This result
is very robust at small scales ($r<$20 Mpc/h), as the volume average
can be properly performed, and it becomes progressively weaker when
the limits of the sample (set by the radius $r_s^m$ of the largest
sphere fully contained in it) are reached. Systematic noise, due to
un-averaged large fluctuations, increases when $r \rightarrow r_s^m$:
one way to overcome this problem is to consider larger samples. In
this respect it is useful to compare our results with the ones derived
by Hogg et al. (2005) by analyzing the largest sample ever studied for
this correlation analysis. In fact they considered a sample of
luminous red galaxies, covering a volume of about $\sim 0.6$
(Gpc/h)$^3$. They found the same power-law as we find here up to 20/30
Mpc/h. They then detected a slow crossover toward homogeneity which is
eventually reached at 70 Mpc/h. With the data we have considered here,
due to the limited solid angle of the survey, we are not able to
confirm or disprove this result. In this respect it is worth noticing
that, for example, Sylos Labini et al. (1998) found a similar value
for the redshift space correlation exponent for the conditional
density at those scales: extending the analysis to larger scales, with
statistical tests of weaker robustness, they however found evidences
for a continuation of correlations with almost the same exponent up to
scales of order of one hundred Mpc/h. Apparently the results by Hogg
et al. (2005) do not confirm such findings.

Leaving the question of the extension of the power-law behavior to
further studies, we focus now on the interpretation of small-scale
correlations. Up to the scale of few tens Mpc/h, the conditional
density $\Gamma(r)$ show a power-law behavior, with exponent
$\gamma=0.8 \pm 0.2$ and well defined amplitude, although with some
fluctuations in different sky regions. As discussed, the amplitude of
the conditional density varies in different VL samples according to
the luminosity of the galaxies selected. This has a very simple
explanation, that brighter galaxies are less frequent than fainter
ones. One can develop an analytical formalism by considering the
effect of the galaxy luminosity function to understand this change: in
the hypothesis that space and luminosity are not correlated, usually
adopted in studies of large scale galaxy distribution, one can
quantitatively compute the amplitude of the conditional density in
different samples.

We have discussed that the results we get for the reduced two-point
correlation function, although in agreement with the ones obtained by
other groups, are affected by finite size effects. The reason is
simply that as long as the distribution presents strong fluctuations,
the study of $\xi(r)$ is problematic. The regime of strong
fluctuations is described by a certain functional behavior of the
conditional density $\Gamma(r)$, in the present case a power-law
function. In this situation the estimation of the sample density is
not only affected by large (statistical) noise, but it becomes sample
size dependent, i.e. by a systematic effect. However because of the
intrinsic large fluctuations systematic and statistical noise are
entangled into the information provided by the amplitude of $\xi(r)$.
Thus explicit tests for systematic finite size effects are needed, and
these are provided by the analysis of the conditional density.

{}

\end{document}